\documentstyle[preprint,prb,aps]{revtex}
\tighten
\begin{document}
%\frontmatter{
\title{Itinerant metamagnetism and possible spin transition in LaCoO$_3$ by temperature/hole doping}
\author{P.~Ravindran$^1$~\cite{byline},H.~Fjellv{\aa}g$^1$,A.~Kjekshus$^1$,
P.~Blaha$^2$, K.~Schwarz$^2$, and J.~Luitz$^2$}
\address{$^1$ Department of Chemistry, University of Oslo, Box 1033, Blindern,
N-0315, Oslo, Norway\\
$^2$Institut f\"{u}r Physikalische und Theoretische Chemie, Technische Universit\"{a}t Wien, 
A-1060 Vienna, Getreidemarkt 9/156, Austria\\
}
\date{\today}
\maketitle
\begin{abstract}
{ The electronic structure of the perovskite La$_{1-x}$Sr$_x$CoO$_3$ has been
obtained as a function of Sr substitution and volume from a series of
generalized-gradient-corrected, full-potential, spin-density-functional
band structure calculations.  The energetics of different spin
configurations are estimated using the fixed-spin-moment (FSM)
method. From the total energy vs spin magnetic moment curve for
LaCoO$_3$ the ground state is found to be nonmagnetic with the Co ions
in a low-spin (LS) state, a result that is consistent with the
experimental observations. Somewhat higher in energy, we find an
intermediate-spin (IS) state with spin moment $\sim$1.2\,$\mu_{B}$/f.u.
From the anomalous temperature dependent susceptibility along with the
observation of an IS state we predict metamagnetism in LaCoO$_3$
originating from an LS-to-IS transition.  The IS state is found to be
metallic and the high-spin (HS) state of LaCoO$_3$ is predicted to be
a half-metallic ferromagnet. With increasing temperature, which is
simulated by a corresponding change of the lattice parameters we have
observed the disappearence of the metamagnetic solution that is
associated with the IS state.  The FSM calculations on
La$_{1-x}$Sr$_x$CoO$_3$ suggest that the hole doping stabilizes the IS
state and the calculated magnetic moments are in good agreement with
the corresponding experimental values.  Our calculations show that the
HS state cannot be stabilized by temperature or hole
doping since the HS state is significantly higher in
energy than the LS or IS state. Hence the spin-state transition in
LaCoO$_3$ by temperature/hole doping is from an LS to an IS spin state
and the present work rules out the other possibilities reported in the
literature.  }
\end{abstract}

%\newpage
\section{INTRODUCTION}
\label{sec:intro}

The coupling of the charge to the spin and lattice degrees of freedom
yields interesting phenomena such as the colossal
magnetoresistance\cite{jin94} and high-temperature
superconductivity\cite{bednorz86} where the underlying mechanism is
still under investigation.  In the La$_{1-x}$Sr$_x$CoO$_3$
perovskite phase , cobalt spin configurations change with temperature and Sr
concentration giving a rich variety of magnetic and transport
properties that has attracted considerable attention over the last
four decades.  The simultaneous presence of strong electron$-$electron
interaction within the transition-metal cobalt $3d$ manifold and a sizable
hopping-interaction strength between the $3d$ and oxygen $2p$ states
are primarily responsible for the wide range of properties exhibited by
these compounds. LaCoO$_3$ is unique in that it is a diamagnetic
semiconductor\cite{yamaguchi96} with a spin gap of 30\,meV, a charge
gap of 0.1\,eV and a rhombohedrally distorted perovskite structure at
low temperature.  It undergoes a spin-state transition from a
diamagnetic state to a paramagnetic state with a finite moment at 100\,K
and from semiconductor to metal above $\sim
500\,K$.\cite{yamaguchi96} To account for the experimentally observed
spin-state transition, both phenomenological explanations and {\em ab initio}
based theories were proposed.

\par 

The magnetic properties of the cobaltites depend on the spin state of
Co$^{3+}$ and Co$^{4+}$, i.e. whether they are in the low-, intermediate-
or high-spin state.  LaCoO$_3$ itself is a nonmagnetic insulator at
low temperature, usually referred to as a low-spin state (LS, $S$ = 0)
because the atomic configuration ($t_{2g}^{6} e_{g}^{0}$) of Co$^{3+}$
ions has no magnetic moment.  Magnetic susceptibility slowly increases
with temperature and reaches a maximum at $T \approx$ 90\,K.  Above
this temperature, the system shows a Curie$-$Weiss$-$law behavior, which is
followed by a structural transition at 500\,K.  The origin of the
low-temperature increase in the susceptibility is unclear at
present. It is suggested that this could be due to frozen-in
ferromagnetic domains which are trapped at low
temperatures.\cite{louca98} Although the spin state of the Co ions and
the nature of the transitions have been investigated for over 40 years
these issues remain controversial.  While it is generally agreed that
the low-temperature phase is a nonmagnetic LS state both a high-spin
state\cite{heikes64,jonker66,raccah67,asai94,abbate93,barman94} (HS,
$S$ = 2, $t_{2g}^{4}e_{g}^{2}$) and an intermediate-spin
state\cite{korotin96,saitoh97} (IS, $S$ = 1, $t_{2g}^{5} e_{g}^{1}$) have
been proposed for LaCoO$_3$ at high temperature.  Most previous
studies have treated the spin-state transition at 90\,K as a
transition from the LS state to a thermally excited HS
state,\cite{raccah67} which is reported\cite{asai89} to be only
10$-$80\,meV higher in energy than the low-spin state.

\par 
The attempts to explain the spin-state transition in LaCoO$_3$ based
on different experimental techniques is rather controversial.  Direct magnetic
measurements allow one to unambiguously identify the LS state in the
temperature range below
50\,K.\cite{jonker66,yamaguchi96,naiman65,rodriguez951,itoh951,saitoh97,asai97}
The behavior of the magnetic susceptibility at low temperature, in the
region around 100\,K has been understood in terms of thermal
activation of a magnetic HS state from the LS ground
state.\cite{jonker66,naiman65,marx80} An anomalous thermal expansion
associated with the spin-state transition has also been
observed\cite{asai94} below 100\,K.  Raccah and
Goodenough\cite{raccah67} have emphasized that more complex supercell
structures play an important role in the spin-state transition and
reported the co-existence of LS and HS states of cobalt in LaCoO$_3$
and a first-order phase change at 1210\,K.  Rodriguez and
Goodenough\cite{rodriguez95} have suggested from magnetic and
transport measurements that LaCoO$_3$ is in an LS state below 35\,K,
an LS-HS disordered state between 35 and 110\,K, an LS-HS ordered
state between 110 and 350\,K and an IS-HS ordered state 
above 650\,K.  However, a careful powder neutron
diffraction study\cite{thornton86} found no evidence for any static
ordering of distinct Co sites.  Abbate {\em et al.} have interpreted
the transition in the range 400$-$650\,K as due to an LS-to-HS
transition, based on x-ray absorption and x-ray photoelectron
spectroscopy measurements.\cite{abbate93} Saitoh {\em et
al.}\cite{saitoh97} have argued that the 100\,K transition is most
likely due to an LS-to-IS transition (from analysis of 
photoemission spectra and magnetic susceptibility data),
and also suggested that the 500\,K transition is due to the
population of the HS state.  The XPS spectra of LaCoO$_3$ at
room temperature and 573\,K along with ionic multiplet analysis
suggest\cite{veal78} that the LS and HS states co-exist at room
temperatures.  Polarized neutron scattering measurements\cite{asai94}
exhibit two magnetic-electronic transitions, one near 90\,K and 
another near 500\,K.  The spin-state transition is maintained to occur
at low temperature, and the high temperature transition is not
dominantly of magnetic origin.  NMR\cite{itoh95} studies also claim
that the LS-to-HS spin state transition occurs at $\sim$ 90\,K.
Co-K-extended x-ray absorption fine structure
measurements\cite{arunarkavalli93} established the occurrence of two
cobalt sites above 400\,K and this may be associated with two spin
states for Co.  Madhusudan {\em et al.}\cite{madhusudan80}
reported that the series RCoO$_3$ (R = Pr, Nd, Tb, Dy and Yb), all
exhibit the considered LS-to-HS state transition of cobalt.
Taguchi\cite{taguchi96} has shown from structural and susceptibility
studies that Co$^{3+}$ in Nd(Cr$_{1-x}$Co$_{x}$)O$_{3}$ is in the 
low-spin state at low temperature and transform to a mixed-spin state with
increasing temperature.  From temperature-dependent susceptibility and
Knight-shift measurements\cite{itoh97} it has been pointed out that
the LS-HS model is not applicable for understanding the spin-state
transtion in LaCoO$_3$ and thus a LS-IS model is more appropriate.
Heikes {\em et al.}\cite{heikes64} considered an IS state to account
for the effective moment obtained from susceptibility data below the
$\sim 500\,K$ transition in LaCoO$_3$.  From photoemission
measurements it has been concluded that the hybridization of oxygen
$2p$ orbitals and cobalt $3d$ states stabilize the IS state with spin
$S$ = 1.\cite{potze95} Heat capacity measurements also support the
LS-to-IS state transition.\cite{stolen97} Recent magnetic
susceptibility and neutron-diffraction studies\cite{caciuffo99} show
that LaCoO$_3$ has the LS Co$^{3+}$ configuration at the lowest
temperatures. Below 350\,K IS remains isolated and localized;
above 650\,K, all the trivalent Co ions are transformed to the IS state
with itinerant $d$ electrons.

\par 
Several theoretical attempts have also been made to understand
the microscopic origin of the spin-state transition in LaCoO$_3$.
From an LDA+U approach (where U is the on-site Coulomb interaction)
Korotin {\em et al.}\cite{korotin96} have demonstrated that the IS
state is relatively stable over the HS state, as a result of the
strong $p-d$ hybridization effect as well as of the orbital ordering
effect. This is in contrast to the expectation from the simple ionic
model.  They have also explained the semiconducting behavior above the
100-to-500\,K region as a spatial ordering of orbitals
associated with a Jahn$-$Teller distortion of an IS state and the
thermal disordering of this state results in a gradual crossover to
the metallic state at high temperatures. However, a controversy still
exists since the calculation of Mizokawa and Fujimori\cite{mizokawa95}
did not find the orbital-ordered state.  Recent neutron diffraction
measurements\cite{asai98} also give no evidence for orbital ordering.
The Hartree$-$Fock calculations on the multi-band lattice
model\cite{mizokawa95,takahashi97} have also shown that the IS state
is more stable than the HS state.  Simulations using
molecular-orbital dynamics\cite{takahashi96} show that magnetoelastic
coupling plays a very important role, the spin-state transition being
mainly induced by variation of the Co$-$O bond length with
temperature.  Liu {\em et al.}\cite{liu93} have suggested that there is
a thermodynamic equilibrium between LS Co$^{3+}$, HS Co$^{3+}$,
Co$^{2+}$ ($t_{2g}^{6}e_{g}^{1}$) and Co$^{4+}$
($t_{2g}^{5}e_{g}^{0}$) in LaCoO$_3$.  The recent
calculations\cite{zhuang98} within the unrestricted Hartree$-$Fock
approximation and a real space recursion method suggest that the
spin-state transition at 90\,K takes place from the LS to LS-HS
ordered state.  Mizokawa and Fujimori\cite{mizokawa96} showed from
unrestricted Hartree$-$Fock calculations that the LS state is
the ground state and the IS state is the first excited spin state.
The Hartree$-$Fock calculation\cite{mizokawa96} also shows that the
total energy of the IS state can be considerably lowered on ordering
of the $e_{g}$ orbitals.  The analysis of the core-level spectrum in
terms of a configuration interaction model suggests that both LS and
HS states coexist at low temperature (100\,K), but at 573\,K there is a
decrease in the LS contribution related to local structural
changes.\cite{barmen94}

\par 
Much of the difficulty in reaching a consensual interpretation
originates from the traditional use of an ionic, ligand-field
model. In such a picture, only two distinct spin states of Co$^{3+}$
are energetically operational.\cite{tanabe54} Also most of the
calculations presented in literature dealing with spin-state
transition involve only a few adjustable parameters.  In order to
obtain a better understanding of both, the spin-state transition in
LaCoO$_3$ and the origin of the paramagnetic state with local magnetic
moments, it is necessary to examine the energetics of various
spin-ordered states as a function of hole doping/temperature. This is
the motivation for the present study.

\par 
The discovery of colossal magnetoresistance (CMR) in 
manganites with perovskite structure\cite{vonhelmolt93} has stimulated
research of compounds exhibiting large magnetoresistance.  Fairly
large magnetoresistance has indeed been observed in the perovskite
series La$_{1-x}$A$_x$CoO$_3$ (A = Ca, Sr or Ba).\cite{briceno95} The
magnetic and transport properties of La$_{1-x}$Sr$_x$CoO$_3$ and CMR
materials such as La$_{1-x}$Sr$_x$MnO$_3$ have common
features.\cite{raccah67,thornton88,bahadur79,itoh94,rodriguez95} In
both system the substitution of La with a divalent ion creates a
metallic ferromagnetic state.  The ferromagnetic interactions between
Co$^{3+}$ and Co$^{4+}$ are supposed to arise from a double exchange
mechanism like that in La$_{1-x}$Sr$_x$MnO$_3$ materials. However, the
detailed mechanism of the ferromagnetism and the metal-insulator
transition in cobaltites are not well understood. In
metallic samples of La$_{1-x}$Sr$_x$CoO$_3$ ($x > 0.2$) the
magnitude of the magnetoresistence is typically small whereas it
becomes larger in the composition range $x \leq 0.2$, where the system
is close to a metal$-$insulator transition.\cite{mahendiran95} 
\par
La$_{1-x}$A$_x$CoO$_3$ are of considerable interest because of the
peculiar way their magnetic and transport properties change with
composition and
temperature.\cite{raccah67,thornton88,bahadur79,itoh94,rodriguez95}
Extensive studies on electronic and magnetic properties of Sr-substituted 
lanthanum cobaltite, La$_{1-x}$Sr$_x$CoO$_3$ have provided
many interesting results during the past four or five decades.  The
magnetic state of La$_{1-x}$A$_x$CoO$_3$ strongly depends on the
doping concentration $x$.  Jonker and van Santen\cite{jonker53} 
studied La$_{1-x}$Sr$_x$CoO$_3$ already in 1953 by
magnetization measurements and reported ferromagnetic order for an
intermediate Sr concentration. They argued that the
double-exchange Co$^{3+}-$Co$^{4+}$ interaction is responsible for the
ferromagnetism.  Bhide {\em et al.}\cite{bhide75} measured the
temperature dependence of M\"{o}ssbauer spectra for ferromagnetic
La$_{1-x}$Sr$_x$CoO$_3$ ($0 \leq x \leq 0.5$) and reported that
ferromagnetic Sr-rich clusters coexist with paramagnetic La-rich
regions in the same (crystallographic speaking) phase. The $3d$ holes created by
the Sr substitution are itinerant both above and below the Curie
temperature ($T_{C}$) and all the experimental results are explained on
the basis of itinerant-electron ferromagnetism.  Itoh {\em et
al.}\cite{itoh94} studied this system by magnetization measurements at
low fields and obtained a phase diagram with a paramagnetic-to-spin-glass 
transition for $x < 0.18$ and a paramagnetic-to-cluster-glass
transition for $x \geq 0.18$. Neutron diffraction
measurements\cite{sathe99} showed the appearance of magnetic peaks
below $T_{C}$ corresponding to long-range ferromagnetic order for $x
\geq 0.2$.

\par The doping changes the valence states of Co to give a mixture of
Co$^{3+}$ and Co$^{4+}$.  Using the unrestricted Hartree$-$Fock
approximation and a real-space-recursion method, Zhuang {\em et
al.}\cite{zhuang981} calculated different magnetic phases for SrCoO$_3$
and concluded that SrCoO$_3$ is in the IS ($t_{2g}^{4}e_{g}^{1}$)
state.  The observed spin moment in La$_{0.7}$Sr$_{0.3}$CoO$_3$ from
neutron diffraction studies\cite{sathe96} has concluded that the
system will have Co$^{4+}$ in the LS state and a mixed LS and HS
configuration for Co$^{3+}$.  The magnetic state of
La$_{1-x}$A$_{x}$CoO$_3$ was suggested to be a mixture of HS Co$^{3+}$
and LS Co$^{4+}$ ($t_{2g}^{5}e_{g}^{0}$),\cite{taguchi78}
because the measured saturation magnetic moment is only about half of
the moment for HS Co$^{3+}$.\cite{taguchi78,jonker53} From application of the
numerically exact diagonalization method on Co$_2$O$_{11}$ clusters it
has been shown that a coexistence of HS and IS due to strong $p$-$d$
mixing is most plausible in doped cobaltites.\cite{tsutsui99} The
evolution with Sr doping of the magnetic and transport data has been
interpreted\cite{rodriguez95} as follows. For $0 < x < 0.1$, the holes
introduced into the CoO$_{3}$ array stabilize the IS clusters trapped
at Sr$^{2+}$ at low temperatures and the IS
clusters become super-paramagnetic below 
$T_{C}$ = 230\,K. In the range $0.1 \leq x \leq 0.18$, the IS clusters
become larger, each containing several holes, and interactions between
the isolated super-paramagnetic clusters give a spin-glass behavior
below $T_{g} < T_{C}$. A collective freezing of the cluster moments
below $T_{g}$ via frustrated inter-cluster interactions has been
confirmed,\cite{mira97} but whether the large IS clusters really are
rich in Sr$^{2+}$ has not been established. However, our recent
electronic structure calculations on La$_{1-x}$Sr$_x$CoO$_3$ with the
supercell approach suggest that Co ions close to both La
and Sr, are in the IS state. 
%#########
In the present paper we report the energetics
of various spin-states of Co ions as a function of hole doping and
volume in LaCoO$_3$. Based on these results we will consider the 
possible spin-state transition in LaCoO$_3$ on temperature and hole
doping.

\par The rest of this paper is organized as follows. In
Sec.\,\ref{sec:details}, we describe the computational procedure used
in the present calculation.  In Sec.\,\ref{sec:resdis}, we discuss the
electronic structure and magnetic properties of LaCoO$_3$ as a
function of Sr substitution and volume from the results obtained from
VCA and supercell calculations using the fixed-spin-moment method.
Finally in Sec.\,\ref{sec:con} the findings of the present study are
summarized.

\section{Computational details}
\label{sec:details}

\subsection{LAPW calculations}

The present investigation is based on {\em ab initio} electronic
structure calculations derived from spin-polarized,
density-functional theory (DFT). In particular we have applied the
full-potential linearized-augmented plane wave (FPLAPW) method as
embodied in the WIEN97 code\cite{wien} using the scalar-relativistic
version without spin-orbit coupling. The charge density and the
potentials are expanded into lattice harmonics up to $L$ = 6 inside the
spheres and into a Fourier series in the interstitial region. We have
included the local orbitals\cite{singh91} for La $5s, 5p$, Sr $4s,
4p$, Co $3p$ and O $2s$.  The effects of exchange and correlation are
treated within the generalized-gradient-corrected local
spin-density approximation using the parameterization scheme of Perdew
{\em et al.}\cite{pw96} We have carried out test calculations
with different sets of {\bf k}-points and found that the FSM curve
changes considerably with the number of {\bf k}-points. For example,
when we use only 4\,{\bf k}-points in our calculations we obtained the
ferromagnetic state lower in energy than the nonmagnetic state.
To ensure convergence for the Brillouin zone
integration 110\,{\bf k}-points in the irreducible wedge of the first
Brillouin zone (IBZ) were used for the rhombohedral structure and
84\,{\bf k}-points in IBZ for the cubic phase (even half the number
of these {\bf k}-points gave the correct result qualitatively). A 
similar density of
{\bf k}-points was used in the supercell calculations.
Self-consistency was achieved by demanding the convergence of the
total energy to be smaller than 10$^{-5}$\,Ry/cell. This corresponds
to a convergence of the charge to below 10$^{-4}$ electrons/atom. For all
calculations the ratio between the sphere radii used are 1.25, 0.88
and 1.25 for $\frac{R_{La}}{R_{O}}$, $\frac{R_{Co}}{R_{O}}$ and
$\frac{R_{Sr}}{R_O}$, respectively, where $R_{O}$=1.8 Bohr. Since the 
spin densities are well
confined within a radius of about 1.5 Bohr, the resulting magnetic
moments do not depend strongly on variation of the atomic sphere
radii.

\subsection{The fixed spin moment (FSM) method}

Conventional spin-polarized calculations based on DFT allows the
moment to float and the ground state is obtained by minimizing the
energy functional, ($E$), with respect to the charge and magnetization
densities, $\rho(r)$ and $m(r)$, under the constraint of a fixed
number of electrons, $N$. With the variational principle this
corresponds to minimizing the functional
\begin{eqnarray}
F[\rho(r),m(r)] = E[\rho(r),m(r)] - \mu [\int \rho(r)dr - N],
\label{eqn1}
\end{eqnarray}
where $\mu$ is the chemical potential. The minimization gives
$\frac{\delta E}{\delta\rho(r)} = \mu, \frac{\delta E}{\delta m(r)}
=0$ leading to effective one-electron equations which are solved by a
self-consistent procedure that determines the moment which minimizes
the total energy. In cases where two (or more) local minima occur for
different moments, conventional spin-polarized calculations become
difficult to converge or "accidently" converge to different
solutions. Although LaCoO$_3$ is a nonmagnetic material, our
conventional spin-polarized calculations always converged to a
ferromagnetic solution with a moment of 1.2\,$\mu_{B}$/f.u.

\par In order to reach convergence or to study spin fluctuations it is
advantageous to calculate the total energy as a function of magnetic
moment using the so-called fixed-spin-moment
method.\cite{fsm,schwarz84,moruzzi86} In this method one uses the magnetic
moment ($M$) as an external parameter and calculates the total energy
as a function of $M$.  In general one must release the constraint that
the Fermi levels for up and down spins are equal because
the equilibrium condition is not satisfied for arbitrary $M$. The
number of valence electrons ($N$) is known and $M$ is fixed (as input
parameter) determining the values for $E_{F}^{\uparrow}$ and
$E_{F}^{\downarrow}$ from\\
\begin{eqnarray}
N = N^{\uparrow} + N^{\downarrow} = \int_{-\infty}^{E_{F}^{\uparrow}}
\label{eqn:nm1}
D^{\uparrow}(E)dE+\int_{-\infty}^{E_{F}^{\downarrow}}D^{\downarrow}(E)dE\\
M = N^{\uparrow}-N^{\downarrow},
\label{eqn:nm2}
\end{eqnarray}
where $D^{\uparrow}(E)$ and $D^{\downarrow}(E)$ are the spin-up and
spin-down DOS, respectively.  At maxima and minima of the FSM curve,
the two Fermi levels are equal,  i.e.  $E_{F}^{\uparrow}$ =
$E_{F}^{\downarrow}$.  In the FSM method Eqn.\ref{eqn1} is modified to
minimize the functional
\begin{eqnarray}
F[\rho(r),m(r)] = E[\rho(r),m(r)] - \mu[\int\rho(r)dr - N] - h[\int m(r)dr -
M]
\label{eqn2}
\end{eqnarray}
where $h$ is the Lagrange multiplier which applies the constraint of
having a fixed spin moment $M$. The orbital contributions to the
magnetization are neglected, since they are usually small for $3d$
transition-metal phases. Instead of minimizing with respect to
$\rho(r)$ and $m(r)$, it is more illustrative to change to the
spin-up, ${\rho^{\uparrow}(r) = \frac{1}{2}[\rho(r) + m(r)]}$, and
spin-down, ${\rho^{\downarrow}(r) = \frac{1}{2}[\rho(r)-m(r)]}$,
densities. Then the variational principle yields $\frac{\delta
E}{\delta\rho^{\uparrow}(r)} = \mu + h$ and $\frac {\delta E}{\delta
\rho^{\downarrow}(r)} = \mu - h$.  $\mu \pm h$ may be identified as
the chemical potentials for the two different spins.  Thus, the
condition of having a fixed spin moment corresponds to using two
different chemical potentials, i.e; two Fermi energies for one material
at zero temperature. This can also be seen by rewriting
of Eqn.\ref{eqn2} as

\begin{eqnarray}
F[\rho(r),m(r)] = E[\rho(r),m(r)] - (\mu+h)[\int\rho^{\uparrow}(r)dr - N^{\uparrow}] 
-(\mu-h)[\int\rho^{\downarrow}(r)dr-N^{\downarrow}].
\label{eqn3}
\end{eqnarray}
Therefore, the FSM method corresponds to fixing the number of
electrons of the two spins separately. Given $N$ and $M$,
$N^{\uparrow}$ and $N^{\downarrow}$ are defined from
Eqns. \ref{eqn:nm1} and \ref{eqn:nm2}. The associated Fermi energies are then found from
the relations
\begin{eqnarray}
N^{\sigma} = \int^{E_{F}^{\sigma}}_{-\infty}D^{\sigma}(E)dE,
\end{eqnarray}
where $D^{\sigma}(E)$ and $E^{\sigma}_{F}$ are the density of states
and the Fermi energy for spin $\sigma$, respectively.  Extensive
FSM calculations have been undertaken to find theoretically the
stability condition of $3d$ and $4d$ magnetic
materials.\cite{moruzzi86,moruzzi88,moruzzi881,krasko87,moruzzi89} 
Similar calculations are also extensively used to explain the
magneto-volume instabilities in Invar alloys.\cite{entel93}

\subsection{Structure}

LaCoO$_3$ has a rhombohedrally distorted pseudo-cubic perovskite
structure.  The space group is $R\bar{3}c$ and each unit cell contains
two formula units.  La$_{1-x}$Sr$_x$CoO$_3$ has the rhombohedral
structure with space group $R\bar{3}c$ in the range $0\leq x \leq 0.5$
and transforms to a cubic phase with the space group $Pm\bar{3}m$ at
higher Sr contents.  The rhombohedral distortion decreases with
increase of $x$. A similar variation results from thermal expansion
for the parent LaCoO$_3$.  The oxygen atoms are in the 6e
($\frac{1}{4}-\delta x,\frac{1}{4}+\delta x,\frac{1}{4}$) Wykoff
position with $\delta x$ =0.0522. Using the force minimization method
we have optimized $\delta x$ and found an optimized value of
0.0638. The FSM calculations at low temperature are made with this value 
for $\delta x$
and the experimental lattice parameters corresponding to 4\,K.  The structural
parameters for the hole-doped systems used in the present calculations
are the same as those used in Ref.\,\onlinecite{ravi99}.  The
spin-state transition is believed to occur mainly due to the larger
variation of the Co$-$O bond length with increasing
temperature.\cite{raccah67,takahashi96} Hence, it is interesting to study
the spin-state transition as a function of lattice parameters and 
we have performed FSM calculations for LaCoO$_3$ with lattice parameters
corresponding to 4\,K as well as 1248\,K.  The structural parameters
at 1248\,K were taken from high-resolution powder-diffraction
measurements.\cite{thornton86}

\subsection{Substitution}

The hole-doping effect in LaCoO$_3$ has been simulated using supercell
calculations as well as virtual-crystal approximation (VCA)
calculations.  For the VCA calculation we have taken into account the
experimentally reported\cite{sathe96} structural parameter changes as
a function of Sr substitution. Hence, these calculations accounted
properly for the hybridization effect. In this approximation the true
atom in the material is replaced by an "average" atom which is
interpolated linearly in charge between the corresponding pure
atoms. In the VCA calculations the charge-transfer effect is not
throughly accounted for, whereas the band-filling effects are 
properly taken into account. The chosen
approximation has an advantage owing to its simplicity and hence we are
able to study small concentrations of Sr in LaCoO$_3$. However, for
50\% Sr substitution we have made explicit supercell FSM calculations.

\section{Results and discussion}
\label{sec:resdis}

Our current results are based on total-energy FPLAPW-band
calculations utilizing the FSM procedure. 
%#### To obtain a physically transparent
%picture as well as for comparison with the results obtained by other experimental
%and theoretical studies we have discussed our findings in terms of an ionic model.
We have considered only the
ferromagnetic state. Each point on the $E(M)$ curves shown in
Fig.\,\ref{fig:fsm4k} is derived from a band calculation in which the
total moment of the phase was constrained to a given $M$ value.
The FSM procedure fixes only the total moment and not the local
moments. The latter are free to take whatever value that minimizes the
energy. In all cases, the local moments vanish at $M$ = 0; i.e., the
situation $M$ = 0 is never produced by a cancellation of opposite local
moments.

\subsection{DOS characteristics}

The calculated DOS for LaCoO$_3$ are shown in Fig.\,\ref{fig:spindos}
for the three spin arrangements LS, IS and HS.  First of all, the
results predict a metallic behavior for the LS state
(Fig. \ref{fig:spindos}c) of LaCoO$_3$, but the total DOS has a sharp
minimum at $E_{F}$.  This is in contrast to the semiconducting behavior
observed experimentally.\cite{thornton88} This kind of discrepancy may
be expected in narrow-band materials due to correlation effect.
However, this does not hold in the present case as our recent optical property
calculations for LaCoO$_3$ show good agreement with experimental
spectra up to 25 eV. Hence, the metallic behavior in LaCoO$_3$ is due
to the usual underestimation of the band gap in LDA and not due to the
correlation effect. We have recently reported\cite{ravi99} on the
bonding behavior of La$_{1-x}$Sr$_x$CoO$_3$ and found indications of
strong covalent hybridization between O $p$ and Co $d$ in this phase.
For a more detailed discussion about the bonding behavior reference
is made to the preceding communication.\cite{ravi99} The O $p$ 
and Co $d$ states are mixed
with each other in the valence band. By contrast, the La spectral
weight contributes mainly to the unoccupied electronic states in the
conduction band. This suggests that La has more ionic character and is
stabilized by the Madelung potential whereas the covalent contribution
to the bonding is more prominent within the Co$-$O octahedra. The
sharp feature at 4\,eV above the Fermi level corresponds to the La
$4f$ band and the DOS segment at 4$-$8\,eV is composed mainly of La
$5d$ states.  
\par The Co $3d$ states are very important and deserve
to be discussed in more detail because they determine the magnetic
properties.  LaCoO$_3$ has a pseudocubic perovskite structure with a
rhombohedral distortion along the (111) direction. Since the
rhombohedral distortion is small, we use the concept of $t_{2g}$ and
$e_{g}$ orbitals (as referred to in the cubic situation) in the
following discussion.  In an octahedral crystal field the $d$
electrons will be split into double degenerate $e_{g}$ and triple
degenerate $t_{2g}$ states.  The $t_{2g}$ and $e_{g}$ projected
density of states of Co in the cubic phases of LaCoO$_3$ and SrCoO$_3$ are
given in Fig.\,\ref{fig:ddos} for the non-spin-polarized and
ferromagnetic cases.  The strong peak below the Fermi level
corresponds to the Co $t_{2g}$ bands. These relatively non-bonding
states produce weak Co$-$O$-$Co interactions which give rise to a
narrow band. By contrast, the Co $e_{g}$ states produce very strong
Co$-$O$-$Co interactions which give rise to a much larger dispersion,
and result in the broader $e_{g}$ band observed for both LaCoO$_3$
and SrCoO$_3$.  When the spin polarization is included in the
calculations, the $t_{2g}$ states for the majority spin become localized
compared with the non-spin-polarized case and thus the total
energy calculation shows that the ferromagnetic state is lower in
energy than in the nonmagnetic case.\cite{ravi99} However, for the
minority-spin state in Fig.\,\ref{fig:ddos}a $E_{F}$ is located at a
sharp non-bonding $t_{2g}$ peak. As this is not a favorable condition
for stability, a rhombohedral distortion arises due to a
Jahn$-$Teller/Peirls-like instability.

\par 
The simple ligand model predicts that for the LS state in
LaCoO$_3$, that the $t_{2g}$ levels are completely filled and the $e_{g}$
states completely empty. In our calculation, however, the $e_{g}$
states are distributed over the entire valence and conduction bands.
The band structure obtained from a tight-binding method\cite{kemp90}
also shows that the $e_{g}$ orbitals partly fall below the Fermi
level and are mixed with the $t_{2g}$ orbitals. This feature can be
understood as follows.  The lobes of the $t_{2g}$ orbitals point
between the oxygen ligands, whereas the $e_{g}$ orbitals point
directly towards the ligands. Hence, the overlap with O $2p$ orbitals will
be greater for the $e_{g}$ states, and the increased overlap results
in local repulsion between overlapping charge densities. This
repulsive interaction pushes the $e_{g}$ orbitals to higher
energy. However, some of the repulsion is compensated by hybridization
in the resulting bonding states which in turn leads to considerable
amount of $e_{g}$ state in the conduction band.  The calculations show
that $E_{F}$ lies in the vicinity of the sharp non-bonding $t_{2g}$ peak
for the Co $3d$ electrons at the ground state LS configuration. In
this case, a small shift in $E_{F}$ leads to large changes in
$N(E_{F})$ and causes the ferromagnetic instability of the system. So,
the presence of the high DOS close to $E_{F}$ and the small energy
difference between the nonmagnetic and ferromagnetic state are the
two main reasons for the temperature induced anomalies in the physical
properties.  

\par 
Considering DOS for LaCoO$_3$ in the rhombohedral structure and the
LS state, it is interesting to note that $E_{F}$ lies in a pseudogap
which separates bonding from anti-bonding states. As a result there is
a gain in one-electron energy and the system stabilizes in the
rhombohedrally distorted nonmagnetic phase.  In the case of SrCoO$_3$,
the $3d$ DOS of Co in the non-spin-polarized case shows that
$E_{F}$ lies in a sharp peak which originates from non-bonding
$t_{2g}$ states. Here the Stoner criterion is fulfilled and the
ferromagnetism appears.  In the spin-polarized case these $t_{2g}$
states are exchange split as shown in Fig.\,\ref{fig:ddos}a.  Unlike
LaCoO$_3$ the Co-$3d$ $t_{2g}$ states for spin-polarized SrCoO$_3$
are well separated from $E_{F}$ and hence stabilizing the ferromagnetic state.
This is evident from the FSM curve for SrCoO$_3$ shown in
Fig.\,\ref{fig:lasrfsm} where the ferromagnetic state with a magnetic
moment 2.6\,$\mu_{B}$/f.u. is 0.463\,eV lower in energy than the
nonmagnetic state.  Even though SrCoO$_3$ is stabilized in the cubic
ferromagnetic phase, our calculation shows that $E_{F}$ falls on a
shoulder of the nonbonding minority spin Co $t_{2g}$ DOS. This may
be one reason for the presence of oxygen vacancies in SrCoO$_3$.  
\par
The pseudo-gap feature in the vicinity of the Fermi level
disappeares in the IS and HS state of LaCoO$_3$ as shown in
Figs.\,\ref{fig:spindos}a and b, respectively.
In Fig.\,\ref{fig:fsm4k} the limits of the IS and HS states are represented by
vertical lines. 
%#############
A detailed examination of energy vs. moment shows that the 
slope change its sign around 1\,$\mu_{B}$. However, a flattened region
is seen in the vicinity of 1.2\,$\mu_{B}$ (Fig.\,\ref{fig:vcafsm} 
and the inset to Fig.\,\ref{fig:fsm4k})
where the ferromagnetic state
gets stabilized according to our ferromagnetic spin-polarized calculations.
So, we assigned a moment of 1.2\,$\mu_{B}$ to the IS state.
From our calculations the IS state turns out to be metallic
which is consistent with the LDA+U calculation \cite{korotin96} that
leads to a half-metallic ferromagnet. Moreover our calculation
predicts the HS state to be a half-metallic ferromagnet whereas 
according to LDA+U calculations it should be a semiconductor. The reason for
the metallic behavior of the IS state is that the bands formed by
$e_{g}$ orbitals are broad and the band splitting is not strong enough
to create a gap. On the other hand, in the HS state, the
$e_{g}^{\uparrow}$ band becomes completely filled and a band gap
appears in the majority-spin state.  DOS for the IS and HS states shows
that the nonbonding $t_{2g}$ electrons are present at $E_{F}$ in the
minority-spin state. Consequently these states are at a energetically
higher level than the LS state (Fig.\,\ref{fig:fsm4k}).  The magnitude
of DOS just below the Fermi level is lower for the IS and HS phases
than for the LS phase. A similar decrease is observed with increasing
values of $x$ in the XPS spectra.\cite{chainani92} In relation to Fig.\,\ref{fig:fsm4k}
it should be noted that the one electron eigenvalue sum
itself explains why the LS state is lower in energy than the IS
and the HS states. 
%######
%Nevertheless we prefer to use an ionic model to explain 
%the findings in
%a physically transparent way as well as for comparison with results
%available in the literature.

\subsection{Hole doping effect}

It has been reported that the La$_{1-x}$Sr$_x$CoO$_3$ phase develops
a ferromagnetic long-range order above $x=0.05$ and that the
metal$-$insulator transition takes place at $x \approx
0.2$.\cite{jonker53,raccah68,bhide75} Optical
measurements\cite{tokura98} show that the electronic structure of the
high-temperature metallic state of LaCoO$_3$ is very similar to that
of the doping-induced metallic state. Hence, the FSM calculations on 
hole-doped LaCoO$_3$ is expected to give a better understanding about the nature of
the temperature-induced spin-state transition.  
\par For the
Sr-doped system, it is natural that the Sr substitution induces a
partial oxidation from Co$^{3+}$ to Co$^{4+}$.  From the effective
magnetic moment obtained by magnetization measurements, Taguchi and
Shimada\cite{taguchi78} concluded that Co$^{3+}$ is in an HS state
in La$_{1-x}$Sr$_x$CoO$_3$
and Co$^{4+}$ in an LS state because the
saturation moment in the ferromagnetic phase ($0.3 > x$) is only about
half the full moment of HS Co$^{3+}$.\cite{jonker53,taguchi78} A
transition from LS to HS through doping has been suggested from NMR
studies on $^{59}$Co and $^{139}$La\cite{itoh95} probes and neutron
scattering.\cite{asai94} Electron spectroscopy\cite{chainani92} and
magnetic measurements\cite{jonker53,raccah68} indicate that hole
doping leads to the formation of HS rather than LS configuration.
As the radius of Sr$^{2+}$ is larger than that of La$^{3+}$, it is
suggested that the Sr doping favors the HS state by the introduction
of Co$^{4+} (3d^{5}$).\cite{munakata97} The HS state of Co$^{3+}$ 
possesses
$S=2$, viz. a maximum magnetic moment of $\mu_{Co}$ =
4\,$\mu_{B}$. This corresponds to the purely ionic model; 
hybridization of Co $3d$ orbitals with the O 2p orbital and the
band formation in the solid states can significantly renormalize this
ionic value.  Hence, the calculated magnetic moments are always smaller
than the ionic values (Table\,\ref{table1}), and
much smaller than the HS ionic value
indicating that this discrepancy cannot be accounted for as a
hybridization effect. It should be noted that one can expect 
a discrepancy between experiment
and theory above $x > 0.5$, where the oxygen deficiency is known to
increase significantly (see the discussion in Ref. 56). 

\par Recent neutron diffraction
studies\cite{caciuffo99} show that the doping with Sr introduces LS
Co$^{4+}$ ($t^{5}e^{0}$) which in turn stabilizes IS Co$^{3+}$ on the
neighbors.  Ferromagnetic resonance measurements suggest that
subjected to hole doping, the cobalt ions transform from a
paramagnetic LS to a ferromagnetic IS state.\cite{bahadur79} Ganguly
{\em et al.}\cite{ganguly94} concluded from magnetic
susceptibility studies that the Co ions in
La$_{0.5}$Sr$_{0.5}$CoO$_{3}$ are in the IS state.  Photo-emission and
x-ray absorption spectroscopic measurements combined with
configuration cluster-model calculations led Saitoh {\em et
al.}\cite{saitoh971} to suggest that it is the Co ions in the IS state
which are responsible for the ferromagnetism in
La$_{1-x}$Sr$_x$CoO$_3$.  If both Co$^{3+}$ and Co$^{4+}$ are in
the IS states in La$_{0.5}$Sr$_{0.5}$CoO$_{3}$, one should expect
2.5\,$\mu_{B}$/f.u. according to an ionic picture.  As in our previous
study\cite{ravi99} (that indicated  strong covalent
bonding between Co and O), we obtained 1.98\,$\mu_{B}$/f.u. for
La$_{0.5}$Sr$_{0.5}$CoO$_{3}$ (Table\,\ref{table1}). The FSM curves
from the VCA calculations (Fig.\,\ref{fig:vcafsm}) as well as the
supercell calculations (Fig.\,\ref{fig:lasrfsm}) give similar
results.  One of the reasons for the stabilization of the IS state by
hole doping is that the hole doping reduces the ionicity and enhances
the covalent hybridization between Co and O.  The observed
stabilization of the IS state in hole-doped LaCoO$_3$ is also
consistent with recent magnetic and transport property
measurements.\cite{kobayashi99} Comparison between the experimental Co
$2p$ x-ray absorption spectra and atomic-multiplet calculations also
indicate that the ground state of SrCoO$_3$ is IS 
($t_{2g}^{4}e_{g}^{1}$).\cite{potze95} For IS Co$^{4+}$ one expects
3\,$\mu_{B}$/atom. But the hybridization effect reduces the magnetic
moment and hence we obtain a value of 2.6\,$\mu_{B}$/f.u. for
SrCoO$_3$.  

\par 
Early studies led to the suggestion that, in the
doped samples, a paramagnetic La$^{3+}$ region co-exists with
ferromagnetic Sr$^{2+}$-rich clusters in the same crystallographic
phase, the ferromagnetic component increasing with $x$.\cite{raccah68}
The magnetic and transport properties of hole-doped LaCoO$_3$
suggest\cite{rodriguez95} that upon Sr doping, the material segregates
into hole-rich, metallic ferromagnetic regions and a hole-poor
matrix similar to LaCoO$_3$. The Co ions of the ferromagnetic phase
are in an IS configurations, the hole-poor region experiences a
thermally induced LS to HS transition. A cluster-glass state in 
the region 0.3 $<$ x $<$ 0.5 is also proposed\cite{itoh94} from 
magnetization measurements. For lower Sr contents ($x <
0.2$), the magnetization measurements have clarified a spin-glass ground
state,\cite{itoh94,asai94} where a strong ferromagnetic short-range
correlation is observed by paramagnetic neutron scattering
experiments.\cite{asai94} Our spin-polarized, supercell calculations
for La$_{0.75}$Sr$_{0.25}$CoO$_{3}$ show that the cobalt ions closer
to Sr and La give significant magnetic contributions. The
calculated magnetic moment for these two kinds of Co ions do not
differ much, indicating that the spin state of these ions are almost
the same and that the hole doping affects almost uniformly all the Co
ions in the structure.  

\par 
From the FSM curve obtained as a function
of $x$ (Fig.\,\ref{fig:vcafsm}), it is clear that the LS phase of
Co$^{3+}$ is the most stable among the magnetic configurations for
$x=0$, but the LS state is found to be unstable with increasing values of
$x$. This is consistent with the results obtained within the HF
approximation.\cite{takahashi97} Further, the instability of the
nonmagnetic phase on hole doping is consistent with the experimentally
observed paramagnetic behavior of the susceptibility for $x \geq
0.08$.\cite{asai94} Hole doping in the LS ground state of pure
LaCoO$_3$ is believed to lead to the formation of localized magnetic
polarons with unusually high spin numbers
($S$ = 10$-$16).\cite{yamaguchi96} For La$_{1-x}$Sr$_x$CoO$_3$, it has
been reported that the spin-state transition around 90\,K disappears
with increasing $x$ and the Co ions are in a magnetic state down to the
lowest temperatures.\cite{raccah68} Our calculated results are
consistent with these observation in the sense that the metastable
state of LaCoO$_3$ disappears and a magnetic phase
appears on hole doping.  

\par
\begin{figure}
\caption{Calculated magnetic moments for La$_{1-x}$Sr$_{x}$CoO$_3$ vs $x$ obtained by 
supercell and VCA calculations from the generalized-gradient-corrected relativistic
full-potential linear muffin-tin orbital method and from FSM calculations  
according to the FPLAPW method. Experimental magnetic moments are taken from
neutron-scattering measurements (Ref. \protect\onlinecite{itoh952}) at 4.2\,K and magnetization
measurements (Ref.\,\protect\onlinecite{rodriguez95}) at 5\,K.}
\label{fig:moment}
\end{figure}
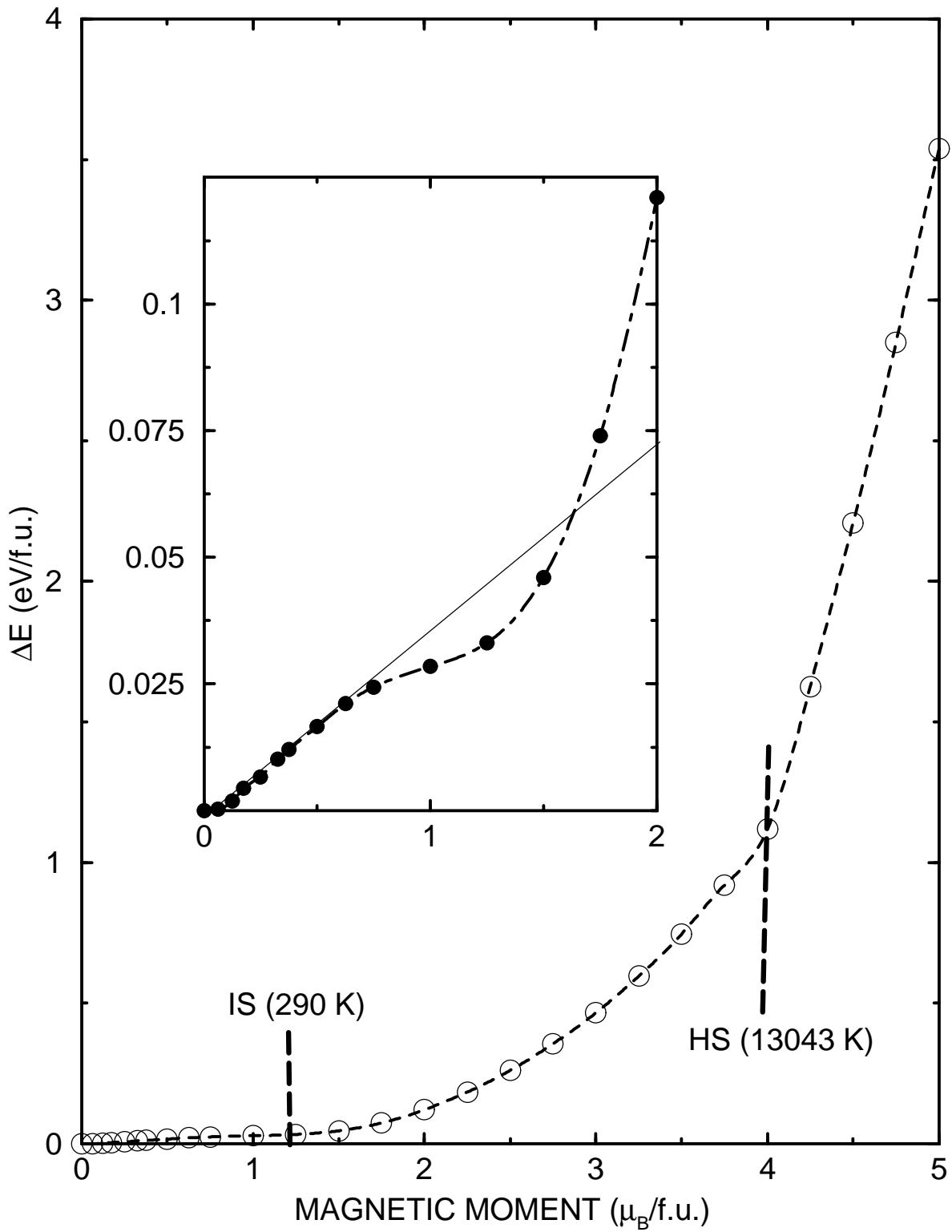
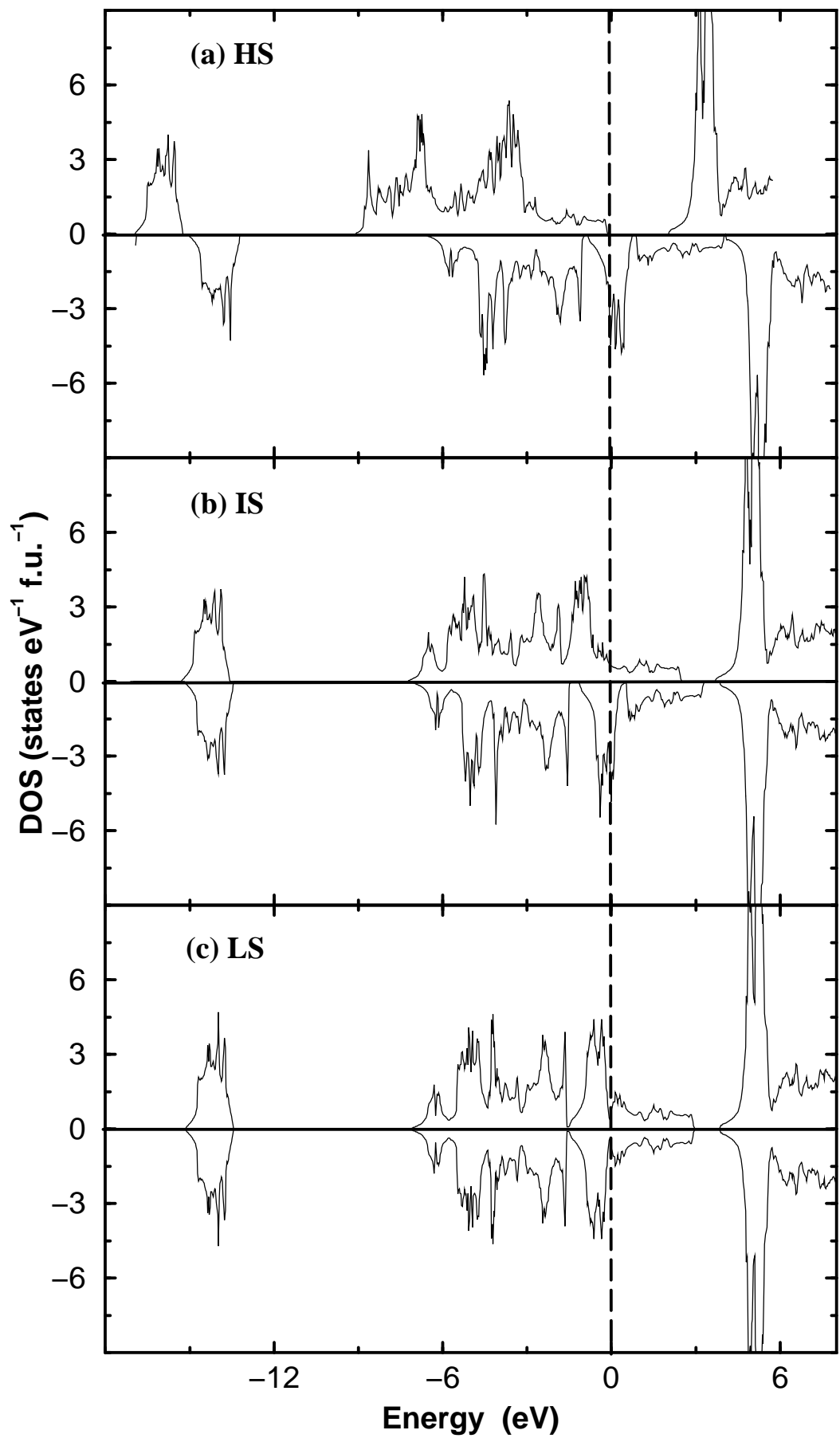
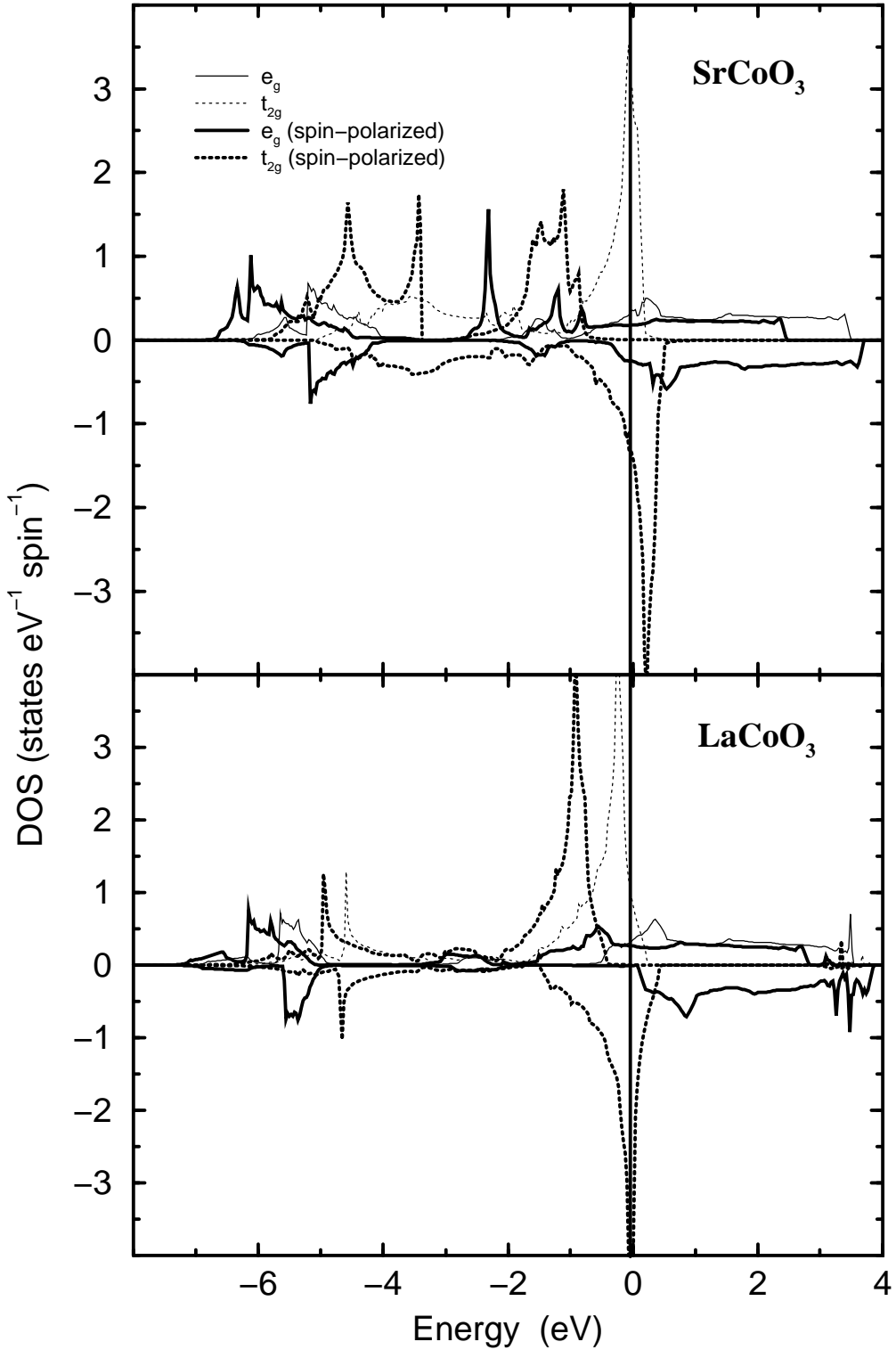
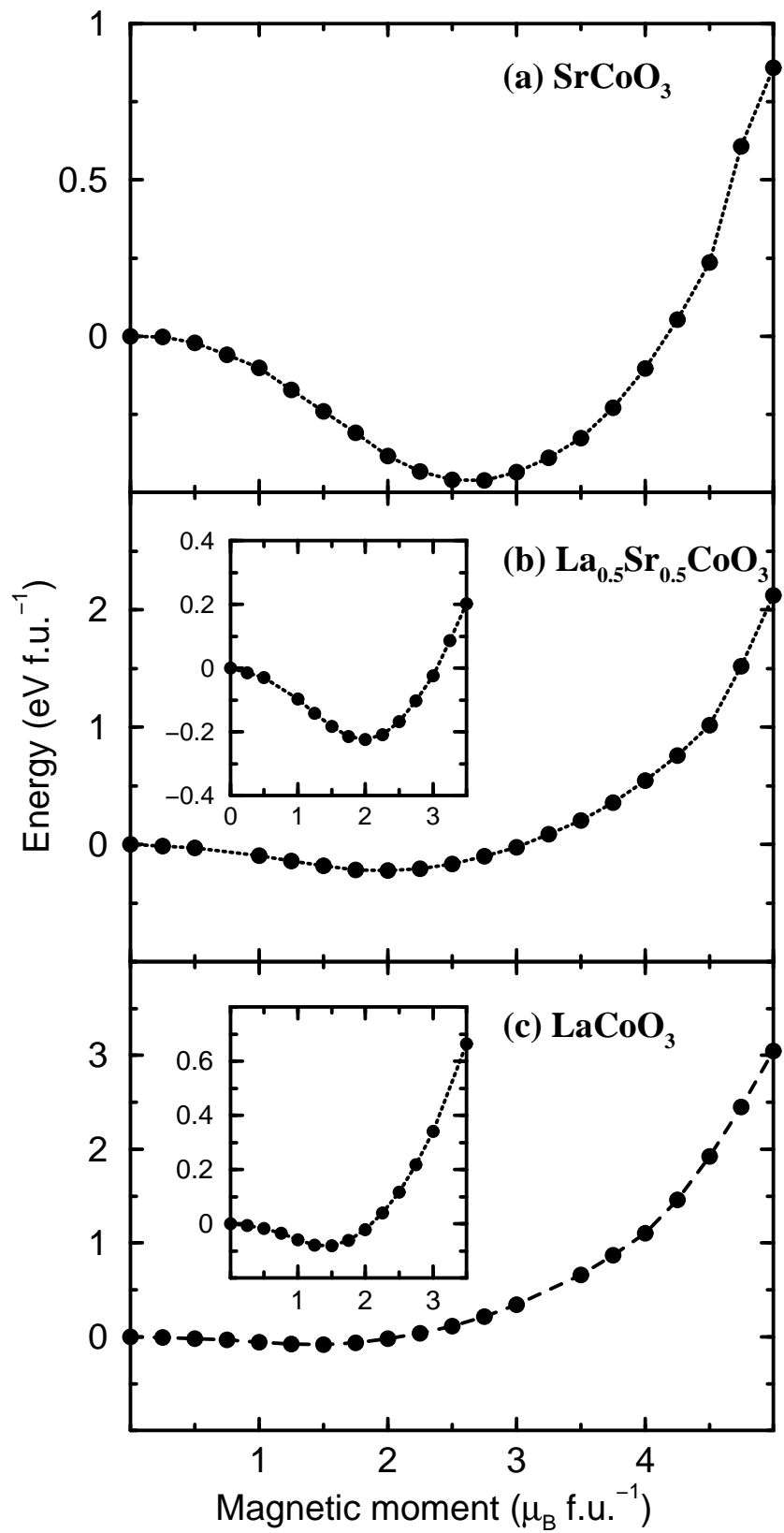
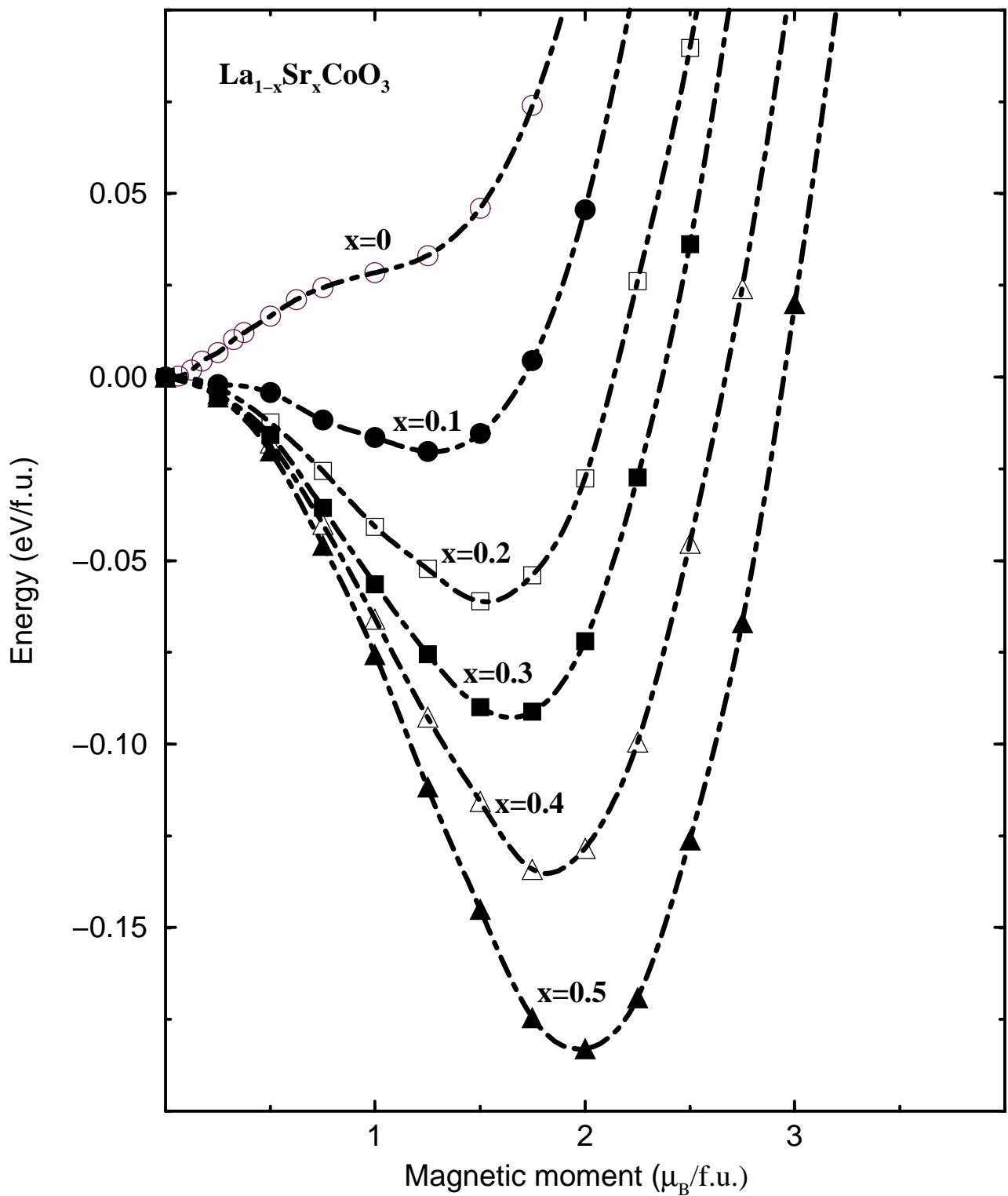
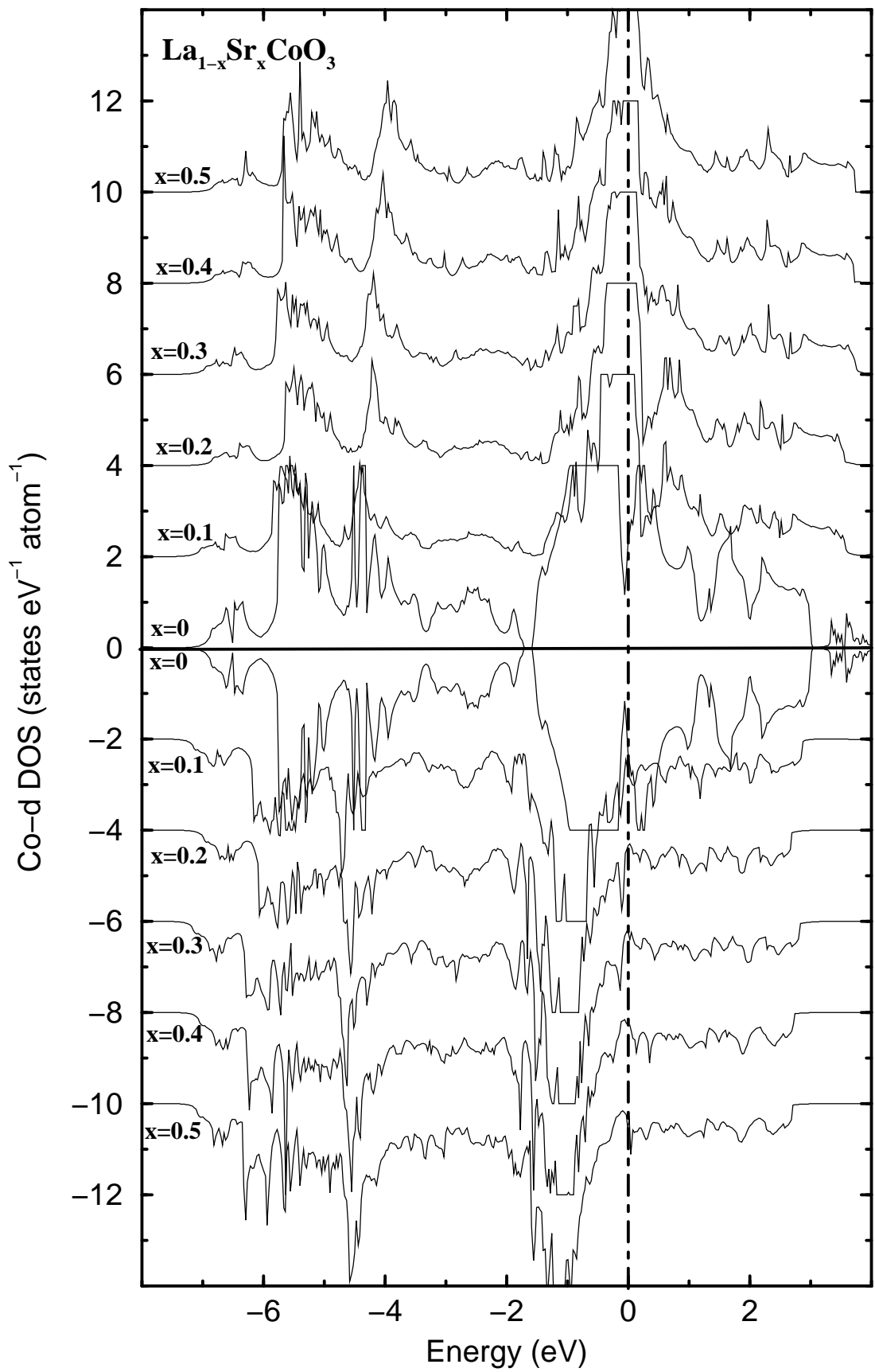
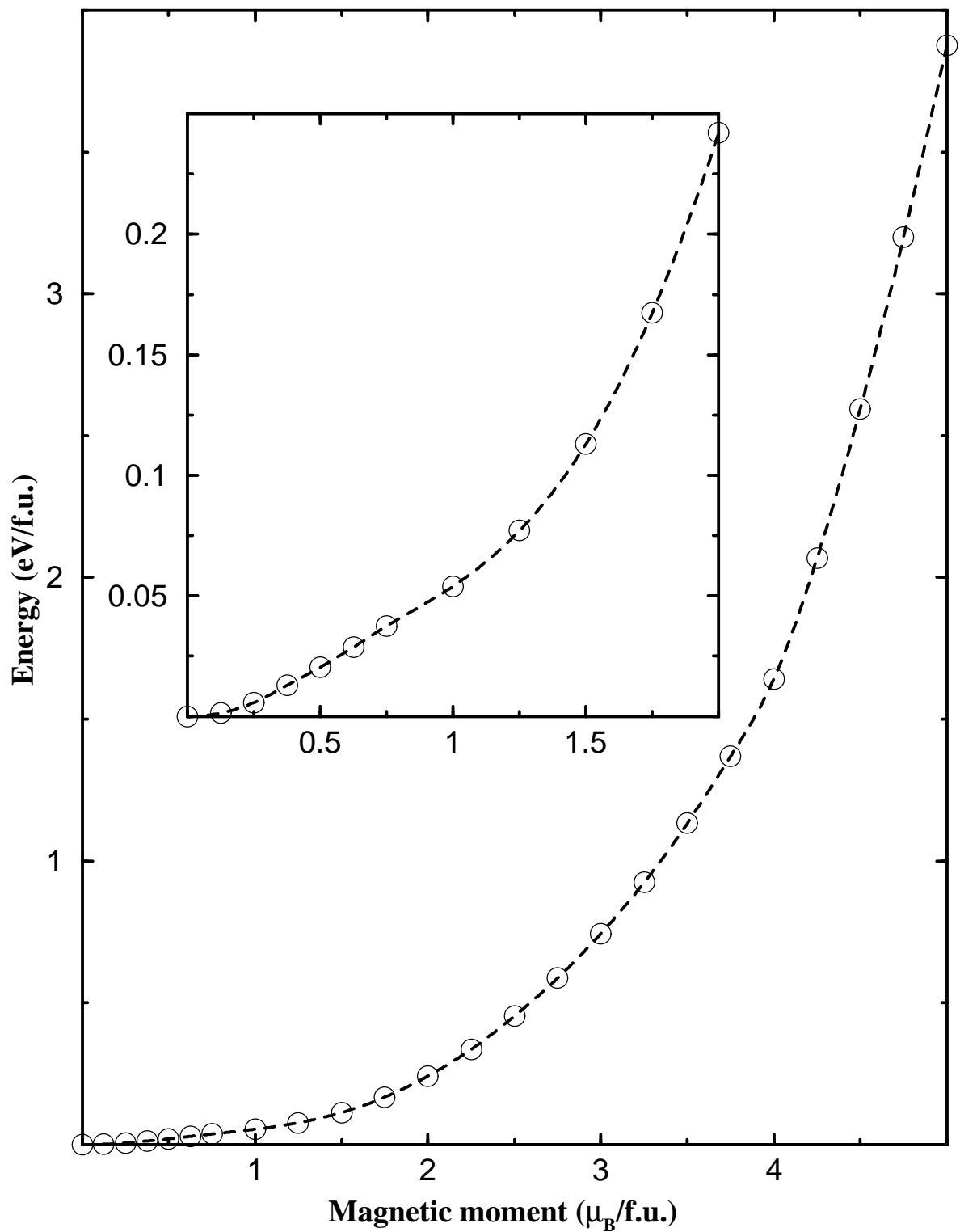
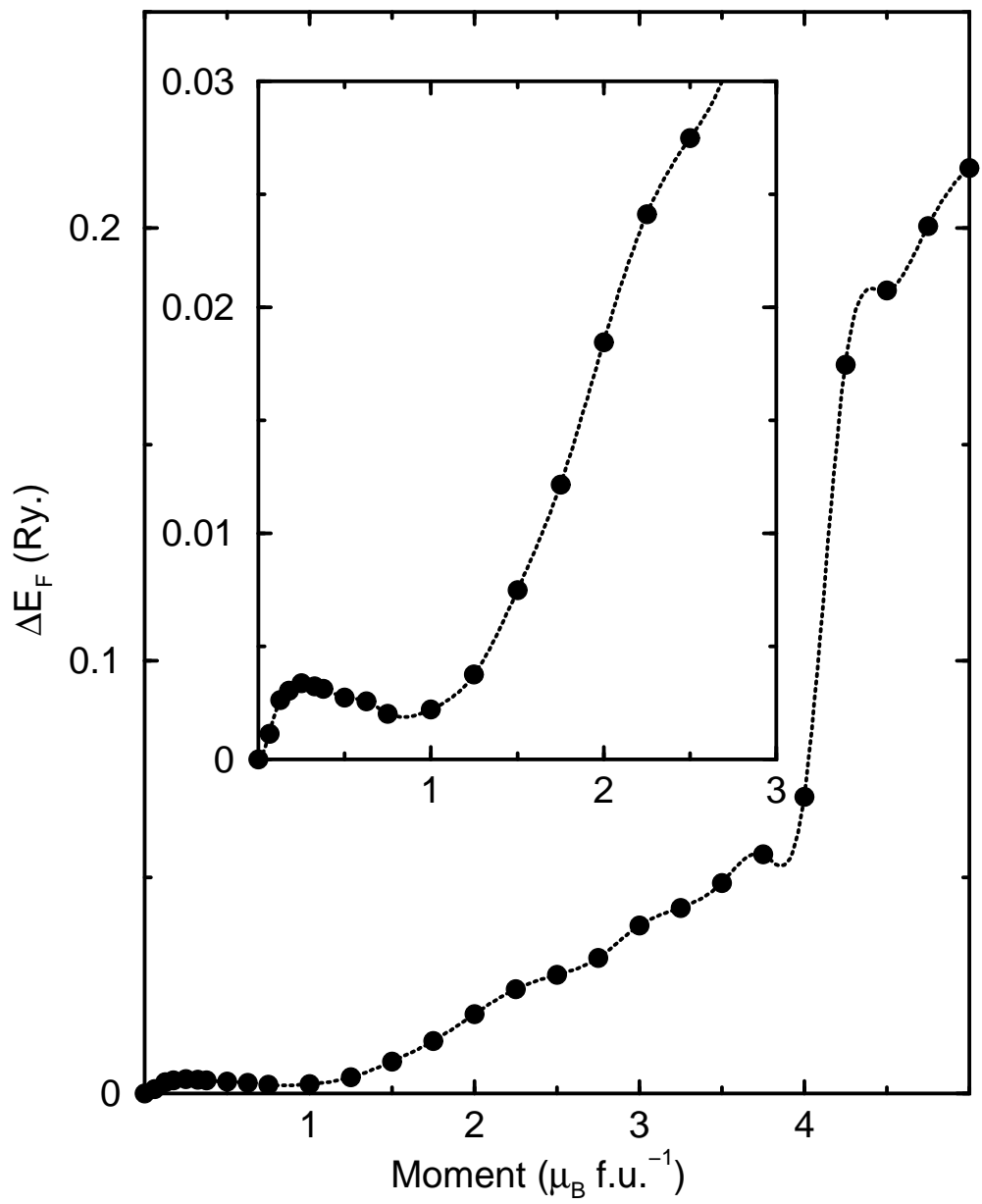
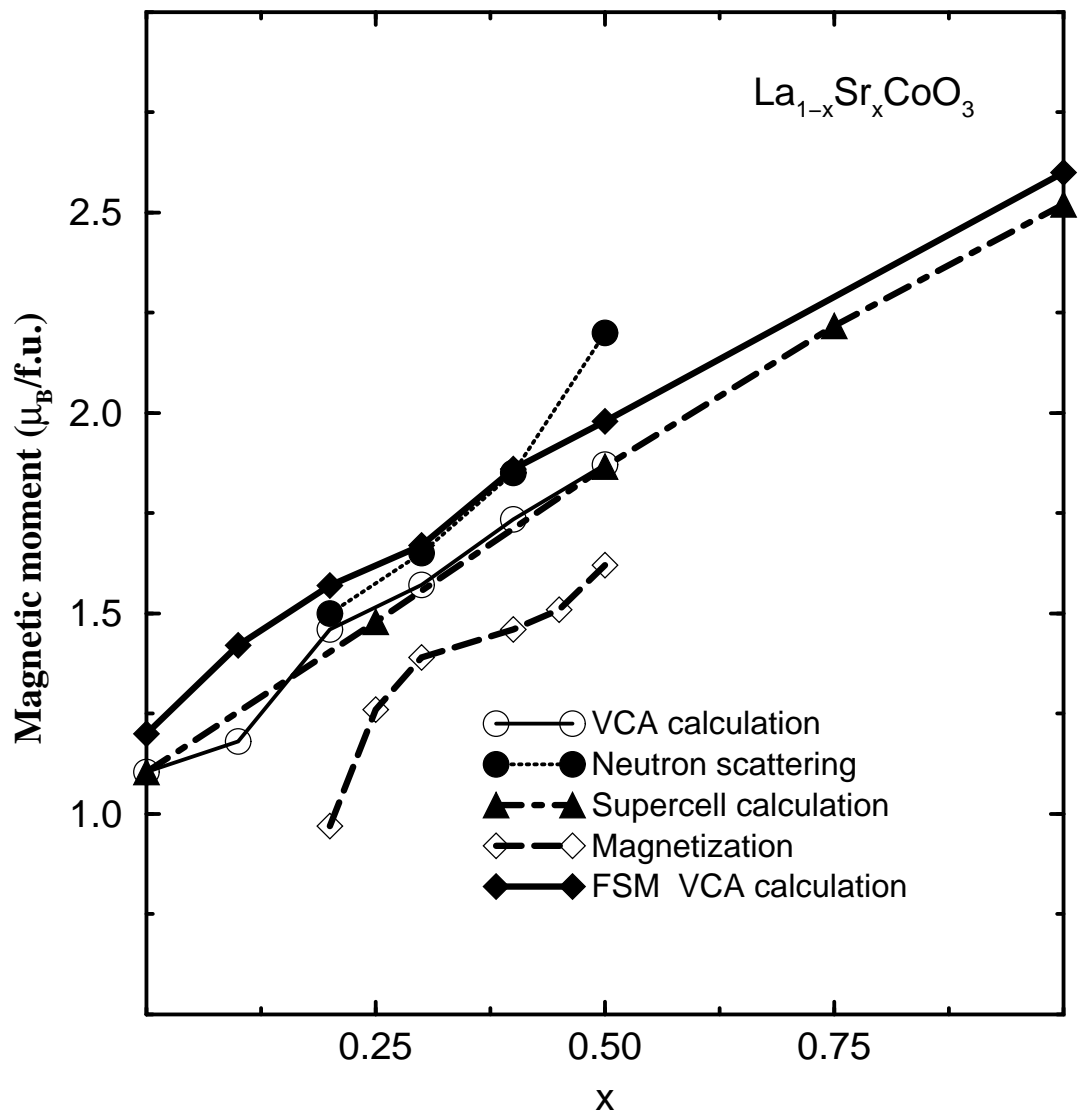

 
The total DOS for LaCoO$_3$ in the LS state (Fig.\,\ref{fig:spindos})
shows that $E_{F}$ is located in a deep valley, viz. in a nonmagnetic
state.  The hole doping shifts $E_{F}$ to the peak on the lower energy
side of DOS (Fig.\,\ref{fig:vcados}).  As a result the Stoner
criterion for band ferromagnetism is fulfilled and magnetism appears.
From Fig.\,\ref{fig:vcados} it is seen that the overall topology
of the DOS curve do not change significantly on hole doping.  
%################
This indicates that the 
band-filling effect plays a decisive role for the changes in the 
magnetic properties of La$_{1-x}$Sr$_x$CoO$_3$ as a function of
$x$.  The equilibrium spin moment
for La$_{1-x}$Sr$_x$CoO$_3$ as a function of $x$, calculated according
to FSM-VCA, is compared with experimental data and the results
obtained by self-consistent full-potential-linear-muffin-tin orbital
(FPLMTO) supercell calculations in Fig.\,\ref{fig:moment}. It should
be noted that the calculated equilibrium magnetic moments seen
from the FSM curves (Fig.\,\ref{fig:vcafsm}) are in excellent
agreement with the experimental findings as well as our previous
FPLMTO results.  This indicates that the FSM method is reliable to
give correct predictions for equilibrium magnetic phases.  The
observation of ferromagnetism in LaCoO$_3$ by temperature/hole doping
has been interpreted as originating from one of the mechanisms: (i) Ordering
of HS and LS Co ions through ferromagnetic super-exchange via the
intervening oxygens,\cite{taguchi78} (ii) Zener double
exchange\cite{wang96} or (iii) itinerant-electron
ferromagnetism.\cite{bhide75,taguchi80} Our band-structure
calculations are able to explain the magnetic properties indicating
that itinerant-band picture (iii) is suitable for these materials. 
%##############
The increasing trend of magnetic moment as a function of hole 
doping originates for two reasons. Firstly, owing 
to band-filling effects the hole doping moves the Fermi
level to the sharp t$_{2g}$ peak and hence enhances DOS 
at the Fermi level as well as the exchange splitting.
Secondly, an effect of reduced hybridization resulting from the
fact that Sr$^{2+}$ is larger in size than La$^{3+}$
and the consequent expansion in lattice on increasing substitution.
As a result the bands narrow which in turn enhances the spin
polarization.

\subsection{Metamagnetism}

A metamagnetic transition refers to paramagnetic metallic materials which are rendered
ferromagnetic by applying a sufficiently large external magnetic
field.  Such a possibility was first discussed by Wohlfarth and
Rhodes\cite{wohlfarth62} and later extended by
Shimizu.\cite{shimizu82} Phenomena related to band metamagnetism are
observed mainly in Co-based materials such as pyrite-type phases
CoS$_2$,\cite{adachi79}, Co(Se,S)$_2$\cite{aachi79}, Laves-type phases
YCo$_2$\cite{goto94}, ScCo$_2$\cite{ishiyama84},
LuCo$_2$\cite{burzo72}, hexagonal Fe$_2$P-type phases Co$_2$P, CoNiP
and orthorhombic Co$_2$P-type\cite{ohta98} phases CoMP (M=Mo and Ru).
Metamagnetism has also been found experimentally\cite{sechovsky86} and
theoretically\cite{eriksson89} in UCoAl. 

\par Itinerant-electron
metamagnets possess an anomalous temperature dependence of the susceptibility
which increases with temperature and then decreases after a maximum at
a finite temperature ($T_{m}$) and usually
obeys the Curie$-$Weiss law at higher temperatures.\cite{goto94,takahashi96} It has been
revealed that $T_{m}$ is closely correlated with the metamagnetic
transition field $H_{m}$.\cite{sakakibara90} A maximum in the
temperature dependence of the paramagnetic susceptibility has been
observed for the itinerant-band metamagnets ScCo$_2$\cite{ishiyama84},
YCo$_2$\cite{lemaire66,burzo72,goto89} and
LuCo$_2$.\cite{burzo72,goto90}  LaCoO$_3$ also shows a temperature
dependent magnetic susceptibility with an increasing trend at
low temperature and a marked maximum around 90\,K followed by a
Curie$-$Weiss-law-like decrease at higher temperatures\cite{asai94,itoh951}
which is similar to that of the itinerant metamagnets.  Metamagnetic
transitions have been recently reported\cite{troyanchuk98} for the closely related phases
Gd$_{0.5}$Ba$_{0.5}$CoO$_{3}$  and
DyCoO$_3$.\cite{schmid73} 

\par The energy needed to enforce a given magnetic moment is given by
the total energy difference $\Delta E$ with respect to the
nonmagnetic case. The $\Delta E$ vs. magnetic-moment curve is shown
in Fig.\,\ref{fig:fsm4k} for LaCoO$_3$ at 4\,K.  At low temperature,
the ground state of LaCoO$_3$ corresponds to the nonmagnetic LS
state. However, the minimum of the ferromagnetic IS state occurs about
32\,meV above the LS state (Fig.\,\ref{fig:fsm4k}).  The existence of
a ferromagnetic solution as a quasi-stable state implies that a
discontinuous transition in magnetization would be possible by
application of a magnetic field. This metamagnetic IS state is
destroyed by temperature as shown in
Fig.\,\ref{fig:fsm1248k}. 
% ###########
In our calculation temperature is introduced via volume
expansion and hence the disappearance of IS by temperature is due to the
weakening of hybridization between Co $d$ and O $p$ states.
%##########
Further, hole doping of LaCoO$_3$ stabilizes
the metastable state (see Fig.\,\ref{fig:vcados}).  The disappearence
of the metamagnetic behavior and the appearence of ferromagnetism on hole
doping in LaCoO$_3$ can be understood as follows. The reduction in
valence electrons by the Sr substitution shifts $E_{F}$ to the lower
energy side of VB towards the peak position in the DOS. As a
result, the Stoner criterion for band ferromagnetism becomes fulfilled
and the metastable state is stabilized as a ferromegnetic state.

\par 
According to the FSM method, the spin-projected DOS
are filled up to $E_{F}$ for spin-up and spin-down electrons in order
to yield the desired externally fixed-spin magnetic moment.  In effect
we are dealing with a system in a uniform magnetic field $H$, which
maintains the magnetic moment of the system. The difference in Fermi
energy corresponds to a difference in magnetic energy
$E_{F}^{\uparrow}- E_{F}^{\downarrow}= 2\,\mu_{B}H$.  Since this
difference is related to the size of the external magnetic field, a
vanishing difference corresponds to an extremum in the magnetic total
energy. Hence, this procedure allows one to find not only stable,
but also metastable magnetic states.  To obtain a better description
of the metamagnetism, the variation of the difference in Fermi
energy for up and down spins ($\Delta E_{F} =
E_{F}^{\uparrow}-E_{F}^{\downarrow}$) with the magnetization, is shown
in Fig.\,\ref{fig:fermi} for LaCoO$_3$ at 4\,K. The difference in
Fermi energy is related to the derivative of $E(M)$ by
\begin{eqnarray}
\Delta E_{F}(M) = -2\mu_{B}\frac{\delta E}{\delta M}.
\end{eqnarray}
Stable zero-field states exist only when $\Delta E_{F}
= 0$ and its derivative, $\frac {\partial\Delta E_{F}}{\partial M}$,
is negative. These solutions then correspond to what one can 
derive by usual spin-polarized calculations.  From 
Fig.\,\ref{fig:fermi} it is seen that $\Delta E_{F}(M)$ nearly
vanishes at the metastable magnetic state obtained in our FSM curve.

\par 
The theoretical exploration of Wohlfarth and Rhodes\cite{wohlfarth62}
revealed that the metamagnetic
transition originates from the sharp peak in DOS in the vicinity of
$E_{F}$. The metamagnetic state in transition-metal
phases is connected with the magnetic-field-induced splitting of the
majority and minority $3d$ subbands of the transition metal.  Bloch
{\em et al.}\cite{bloch75} have pointed out that, according to the
Stoner theory, $E_{F}$ of the itinerant $d$ electrons lies near a
minimum or on a steep decrease in the DOS curve, so that there appears
anomalous temperature and magnetic field dependences in the magnetic
susceptibility. It is interesting to note that the DOS curve for
LaCoO$_3$ in the LS state (Fig.\,\ref{fig:spindos}c) also shows
$E_{F}$ in a deep valley just above the Co $3d$ $t_{2g}$ peak. In this
case the ferromagnetic state with a rather large magnetic moment can
be approached by the applied magnetic field.  This may be the possible
origin for the appearance of the metamagnetism as well as the
anomalous behavior in the temperature dependence of magnetic
susceptibility.  

\par The metamagnetic transition from paramagnetic to
ferromagnetic state can be understood as follows. The applied external
field shifts the up- and down-spin bands by a small amount. As a
consequence the system gains potential energy owing to the reduced
Coulomb interaction, but at the same time loses energy owing to an
increase in the kinetic energy. The final induced magnetic moment will
be determined by the balance between these two terms. However, in some
systems there will be, for a certain critical value of the applied
field, a sudden gain in potential energy which is not sufficiently
counterbalanced by the kinetic term. Hence, the metamagnetic spin state
develops.\cite{cyrot79} More quantitatively, the criterion for the
onset of metamagnetism may be written\cite{cyrot79}
\[
I(\frac{1}{2}D^{\uparrow}(E_{F})+\frac{1}{2}D^{\downarrow}(E_{F}))^{-1} = 1
\]
where $I$ is the multi-band Stoner parameter, and $D^{\uparrow}(E_{F})$ and
$D^{\downarrow}(E_{F})$ are DOS at $N_{F}$ for the spin-up and -down
bands, respectively. If the densities of the two state simultaneously attain 
sufficiently large values for a given applied field,
a metamagnetic transition will take place. A glance at Fig.\,\ref{fig:spindos} shows
that a small splitting of the spin band will indeed give large values of both
$D^{\uparrow}(E_{F})$ and $D^{\downarrow}(E_{F})$. This explains the 
occurrence of the metamagnetic transition in LaCoO$_3$.

\subsection{Spin-state transition}
Despite numerous studies the nature of the spin-state transition is
still under debate. 
This is because in most earlier studies a rather ionic and ligand-field-like starting
point has been assumed. Such a simplified picture does not include the possibility of an
IS state and a large $e_{g}$ band width.
%#####
The main controversies are related to the nature of the
transition, the temperature range of the transition, and to the electronic
structure and hence to the energy levels involved.  From the
temperature dependent $^{59}$Co and $^{139}$La NMR measurements Itoh
and Natori\cite{itoh95} have shown that the anomalous behavior around
500\,K can be interpreted as a spin-state transition between IS and
HS.  From electrical resistivity and neutron diffraction
measurements Thornton {\em et al.}\cite{thornton82} concluded that the
semiconductor-to-metal transition takes place around 520 to 750\,K by
the stabilization of an IS state for Co$^{3+}$ associated
with a smooth transition from localized $e_{g}$ to itinerant
$\sigma^{*}$ electrons.  The polarized neutron-scattering
experiments\cite{asai94} firmly supported that the spin-state
transition takes place at about 90\,K which is in sharp contrast to
the interpretation of x-ray absorption spectroscopy\cite{abbate93}
data which indicates that the spin-state transition takes place in the
range 400$-$650\,K coinciding with the gradual semiconductor-to-metal
transition.  It is also suggested that the spin-state transition in
LaCoO$_3$ occurs in two steps.\cite{asai94,asai98} First, a conversion
from the ground state LS to IS around 100\,K. Second, a
change from the IS state to a mixed IS$-$HS state around 500\,K.  Our
calculations predicts that the energy barrier between the LS and IS
states is 290\,K and that the HS state is much higher in energy than IS
(Fig.\,\ref{fig:fsm4k}).  
%########
So, the possibility of stabilizing the HS
state by temperature and/or hole doping is less favorable in
La$_{1-x}$Sr$_x$CoO$_3$. Hence, the present calculations rule out
the possibility of stabilizing mixed-spin states such as LS-HS and
IS-HS.

\par 
The energy difference between the LS
and HS states of Co$^{3+}$ ions is reported\cite{heikes64} to be less
than 80\,meV and may even be as low as 30\,meV.\cite{yamaguchi96}  From
magnetization measurements the energy difference between the LS
and HS states is reported to be 6 to
22\,meV.\cite{liu93,raccah67,bhide72}  Asai {\em et al.}\cite{asai97}
estimated the energies of the IS and HS states at 0\,K to be 22.5 and
124.6\,meV above LS, respectively.  Our calculated value for the energy
difference between the LS and IS states is 32\,meV and that between LS
and HS states 1113\,meV (see Fig.\,\ref{fig:fsm4k}).  The
unrestricted Hartree$-$Fock calculation shows that the energy of the
HS state is 35\,meV higher than the LS state.\cite{zhuang98} The
Hartree$-$Fock calculation yields the total energy of the IS state
about 0.5\,eV higher than the LS state.\cite{takahashi97} Possible
reasons for the discrepancy between the two sets of results may be
that the Hartree$-$Fock calculations have not taken into account the
structural distortions and all possible low-energy phases. Our
recent electronic structure studies show\cite{ravi99} that the
rhombohedral distortion is important for a correct prediction of the
magnetic properties in LaCoO$_3$ (viz. calculations predicting the cubic
phase to be ferromagnetic). From Fig.\,\ref{fig:lasrfsm}a, the cubic
phase of LaCoO$_3$ is found to be in a ferromagnetic IS state with a
magnetic moment of 1.44\,$\mu_{B}$/f.u. and this state is 82\,meV lower
in energy than the nonmagnetic phase.  

\par 
On going through the
intermediate to high temperature range, LaCoO$_3$ is variously
believed to be in a mixed
LS/HS\cite{heikes64,raccah67,bhide72,madhusudan80,vasanthacharya83,itoh95,rodriguez951,yamaguchi96,asai89,asai94,veal78,barman94},
IS\cite{vasanthacharya83,asai98,potze95,abbate93} mixed
IS/HS\cite{asai98} or
HS\cite{heikes64,raccah67,bhide72,madhusudan80,vasanthacharya83,itoh95,rodriguez951,yamaguchi96,asai89,asai94,veal78,barman94}
state.  Softening of the lattice during the spin-state transition has
also been observed in the elastic modulus and this is explained by a
model involving three spin states coupled with the lattice.\cite{murata99}
Using the LDA+U approach Korotin {\em et al.}\cite{korotin96} found
that the ground state of Co$^{3+}$ in LaCoO$_{3}$ is a LS nonmagnetic
state and found two IS states followed by an HS state at significantly
higher energy.\cite{korotin96} {\em Ab initio} electronic structure
studies on LaCoO$_{3}-$SrCoO$_3$ concluded that on decreasing the
crystal field LS becomes unstable, while the HS state becomes
stabilized.\cite{takahashi98} The LS-to-HS transition model needs
antiferromagnetic interaction between HS Co$^{3+}$ in order to explain
the small absolute value of the magnetic
susceptibility,\cite{yamaguchi96,itoh951} whereas the inelastic
neutron scattering study has revealed the presence of weak
ferromagnetic correlations at $T \geq 100\,K$.  

\par 
Being a $d^{6}$
phase, LS would be more stable than HS if the crystal field
splitting of the $3d$ states into $t_{2g}$ and $e_{g}$ levels is
larger than the intra-atomic exchange splitting (i.e. $10Dq > 2J$).
Conversely, if the exchange energy dominates, the result is an HS state
with $S$ = 2. Owing to the degenerate nature of the IS state with $S$ = 1,
it is not possible to obtain stabilization within this framework.
However, if hybridization with the oxygen band is taken into account
the stability of IS can be accounted for.  In
Fig.\,\ref{fig:fsm4k}, nearly degeneracy of the LS and IS states occurs
because the intra-atomic exchange splitting $\Delta_{ex}$ of the $S$ = 1
state is close to the crystal-field splitting $\Delta_{cf}$.  The
present calculation shows that in the metastable magnetic state, La, Co
and O carry moments of 0, 0.968 and 0.05 $\mu_{B}$/atom, respectively,
for the experimental lattice parameters at 4\,K. These values along
with the contribution from the interstitial region give a total
magnetic moment of 1.2\,$\mu_{B}$/f.u.  

\par It should be noted that
our spin-polarized calculation for LaCoO$_3$\cite{ravi99} always yield
a finite moment corresponding to the flat region currently obtained
from our FSM calculation for 4\,K. From the FSM curve in
Fig.\,\ref{fig:fsm4k} the calculated magnetic field necessary to
stabilize an IS state is 282\,T.  It is well known that Co$^{3+}$ in the HS
state have a larger ionic radius than in the LS state, and LS$-$HS
transitions are believed to be accompanied by an increase of in
volume. Keeping those features in mind we carried out the calculations
for the electronic structure of LaCoO$_3$ with the lattice parameters
for 1248\,K which can imitate the influence of the temperature via
thermal expansion. Fig.\,\ref{fig:fsm1248k} shows the total energies
as a function of magnetic moment for the expanded lattice.  The HS
state ($M$ = $4\,\mu_{B}$) lies much higher in energy than LS and IS at
low as well as high temperature indicating that it is less probable to
stabilize the HS state by temperature/hole doping. However, our FSM
calculations show that an IS state is always lower in energy than the
HS state for hole doped LaCoO$_3$. So, theory rules out the
possibility of stabilizing HS by either temperature or hole doping.
Hence, the LS-to-IS spin-state transition is the more probable
transition in LaCoO$_3$.

\section{SUMMARY}
\label{sec:con}

The electronic structure and magnetic properties of LaCoO$_3$ as a
function of hole doping and temperature have been studied with the
fixed-spin procedure using the generalized-gradient-corrected FLAPW
method.  The hole-doping effect has been simulated by VCA as well as
through supercell calculations. From these studies the following
conclusions have been arrived.\\

1. The nonmagnetic solution with Co$^{3+}$ in the LS state has the
lowest total energy at low temperature and this is consistent with the
experimental observations.\\

2. The FSM calculation suggests that the first excited state
configuration in LaCoO$_3$ is an IS state which lies only 32\,meV
higher than the LS state.  It is experimentally established that
there is a transition from an LS nonmagnetic state to a magnetic state
with increasing temperature. As this transition is experimentally
reported to be around 90\,K, we ascribe the nonmagnetic-to-magnetic
transition in LaCoO$_3$ to the LS$-$IS transition.  
Theoretical results along with the experimental finding lead us to the
conclusion that the transition near 500\,K is not a spin transition,
but rather a semiconductor-to-metal transition and/or a
localized-to-itenerant electron transition.\\

3. The IS state is very sensitive to the volume and it becomes
unfavorable at a volume corresponding to the experimental volume at
1248\,K. \\

4. The HS state is expected to have a moment of 4\,$\mu_{B}$/Co$^{3+}$
within the ionic picture which is much higher in energy than the IS
state at 4\,K as well as at 1248\,K. Hence, the possible spin-state
transition in LaCoO$_3$ is from LS to IS and theory predicts that the
stabilization of the HS state in LaCoO$_3$ is less probable. \\

5. We find that the energy barrier between LS and HS as well as
between IS and HS are very large. 
%####
Also our supercell calculations on
hole doped LaCoO$_3$ show that the Co ions closer to La and Sr 
have almost the same spin polarization. Hence theory rules out the
possibility of stabilizing a mixed-spin state such as LS and HS, IS
and HS or the combination of these by temperature/hole doping. \\

6. The hole doping effect induces ferromagnetism in LaCoO$_3$ which
originates from the filling of Co $t_{2g}$ levels and as a result the
IS state gets stabilized over the LS state. The calculated magnetic
moment for the equilibrium state as a function of hole doping is found
to be in good agreement with the low temperature neutron diffraction
and susceptibility data. \\

7. With equilibrium lattice parameters, our calculations predict the
possibility of a metamagnetic transition.  Hence the experimentally
observed anomalous behavior in the temperature dependent
susceptibility has been interpreted as a metamagnetic transition from
a nonmagnetic LS to a ferromagnetic IS state.  The critical field for
the metamagnetic transition is estimated from the FSM curve of LaCoO$_3$
to be 282\,T.\\

\acknowledgements 
PR is thankful for the financial support from the
Research Council of Norway. Part of these calculations were carried
out on the Norwegian supercomputer facilities. 
PR is also grateful to
%Prof. B. Johansson, Prof. O.  Eriksson and Dr. Lars Nordstr\"{o}m for
%fruitful discussions and 
Mrs. R. Vidya for the comments and critical 
reading of the manuscript.

%%%%%%%%%%%%%%%%%%%%%%%%% Table 1 %%%%%%%%%%%%%%%%%%%%%%%%%%%%%%%%%%
\begin{table}
\caption {Magnetic moments in $\mu_{B}$/Co for La$_{1-x}$Sr$_x$CoO$_3$ obtained
by FSM calculations (Theory) compared with experimental
neutron scattering values (Exp.), together with values obtained for
LS, IS and HS configurations using an ionic model.}
\begin{tabular}{l|r|r|r|r|l}
Composition               & LS  &    IS &   HS &  Exp.&  Theory\\
\hline
LaCoO$_3$                 & 0  &  2  &   4  & 0 & 0 \\
La$_{0.9}$Sr$_{0.1}$CoO$_{3}$ & 0.1  & 2.1  &  4.1 & ---  & 1.42\\
La$_{0.8}$Sr$_{0.2}$CoO$_{3}$ & 0.2 &  2.2  &  4.2 & 1.5&  1.57\\
La$_{0.7}$Sr$_{0.3}$CoO$_{3}$  & 0.3 &   2.3 &    4.3 &  1.65 &  1.67\\
La$_{0.6}$Sr$_{0.4}$CoO$_{3}$ & 0.4 &  2.4 &   4.4&  1.85&  1.86\\
La$_{0.5}$Sr$_{0.5}$CoO$_{3}$ & 0.5 &  2.5 &   4.5&  2.2 &  1.98\\
SrCoO$_3$                 &  1  &  3   &  5& 1.8&  2.60\\
\end{tabular}
\label{table1}
\end{table}

%%%%%%%%%%%%%%%%%%%%%%%%% FIG.1 %%%%%%%%%%%%%%%%%%%%%%%%%%%%%%%%%%
\begin{figure}
\caption{
The total energy of LaCoO$_3$ as a function of constrained spin moment
obtained for a volume corresponding to 4\,K. The zero
level is chosen at the nonmagentic energy minimum. The inset gives an
enlargement for the energy interval close to 
IS. The extrapolation of the metastable state above 25\,meV
is indicated by the straight line. Vertical dashed
lines represent the cross-over to IS and HS and the 
corresponding temperatures required
to excite the system to these states are quoted. 
}
\label{fig:fsm4k}
\end{figure}
%%%%%%%%%%%%%%%%%%%%%%%%% FIG.2 %%%%%%%%%%%%%%%%%%%%%%%%%%%%%%%%%%
\begin{figure}
\caption{Total DOS for LaCoO$_3$ in the LS, IS and the HS states
obtained for the volume corresponding to 4\,K.
The Fermi level is marked with the vertical line at zero energy.}
\label{fig:spindos}
\end{figure}
%%%%%%%%%%%%%%%%%%%%%%%%% FIG.3 %%%%%%%%%%%%%%%%%%%%%%%%%%%%%%%%%%
\begin{figure}
\caption{
The $e_{g}$ and $t_{2g}$ splitted Co-$d$ DOS for (a) LaCoO$_3$ and (b) SrCoO$_3$
with and without spin-polarization.}
\label{fig:ddos}
\end{figure}
%%%%%%%%%%%%%%%%%%%%%%%%% FIG.4 %%%%%%%%%%%%%%%%%%%%%%%%%%%%%%%%%%
\begin{figure}
\caption{
The calculated FSM curves for the cubic phase of (a) SrCoO$_3$, (b) La$_{0.5}$Sr$_{0.5}Co$O$_3$
and (c) LaCoO$_3$.}
\label{fig:lasrfsm}
\end{figure}
%%%%%%%%%%%%%%%%%%%%%%%%% FIG.5 %%%%%%%%%%%%%%%%%%%%%%%%%%%%%%%%%%
\begin{figure}
\caption{
FSM curves for La$_{1-x}$Sr$_{x}$CoO$_3$ ($x$ = 0 to 0.5)
obtained by VCA calculation.}
\label{fig:vcafsm}
\end{figure}
%%%%%%%%%%%%%%%%%%%%%%%%% FIG.6 %%%%%%%%%%%%%%%%%%%%%%%%%%%%%%%%%%
\begin{figure}
\caption{Co $d$-DOS in La$_{1-x}$Sr$_{x}$CoO$_3$ as a function of $x$ obtained from spin-polarized
VCA calculation. For clarity the DOS curves are 
systematically
shifted 2 states eV$^{-1}$ atom$^{-1}$ in both spin channels for each increment of 0.1 in $x$,
and the DOS maxima are cut at 4 states eV$^{-1}$ atom$^{-1}$.}
\label{fig:vcados}
\end{figure}
%%%%%%%%%%%%%%%%%%%%%%%%% FIG.7 %%%%%%%%%%%%%%%%%%%%%%%%%%%%%%%%%%
\begin{figure}
\caption{
The total energy for LaCoO$_3$ as a function of constrained spin moment obtained
for a volume corresponding to 1248\,K.}
\label{fig:fsm1248k}
\end{figure}
%%%%%%%%%%%%%%%%%%%%%%%%% FIG.8 %%%%%%%%%%%%%%%%%%%%%%%%%%%%%%%%%%
\begin{figure}
\caption{ The difference in Fermi energy of majority and minority spin
states of LaCoO$_3$ as a function of magnetic moment for the volume corresponding
to 4\,K.}
\label{fig:fermi}
\end{figure}
%%%%%%%%%%%%%%%%%%%%%%%%% FIG.9 %%%%%%%%%%%%%%%%%%%%%%%%%%%%%%%%%%
\end{document}